\documentclass[11pt,a4paper]{article}
\pdfoutput=1
\usepackage{jheppub}
\usepackage[utf8]{inputenc}

\usepackage{color}
\usepackage{amsmath}
\usepackage{mathtools}
\usepackage{esint}
\usepackage{pifont}
\usepackage{bbold}

\usepackage{bbm}
\usepackage{verbatim}   
\usepackage{subfigure}
\usepackage{acronym}

\usepackage{amsfonts}
\usepackage{amssymb}
\usepackage{mathrsfs}
\usepackage{graphicx}
\usepackage{multirow}
 \usepackage{slashed}

\usepackage{caption}

\usepackage{soul}

\definecolor{mypurple}{RGB}{164,64,214}

\title{Rescuing Massive Photons from the Swampland}

\author[a]{Nathaniel Craig}
\emailAdd{ncraig@physics.ucsb.edu}
\author[b]{and Isabel Garc\'ia Garc\'ia}
\emailAdd{isabel@kitp.ucsb.edu}

\affiliation[a]{Department of Physics, University of California, Santa Barbara, CA 93106, USA}
\affiliation[b]{Kavli Institute for Theoretical Physics, University of California, Santa Barbara, CA 93106, USA}

\abstract{Stringent Swampland conjectures aimed at effective theories containing massive abelian vectors have recently been proposed \cite{Reece}, with striking phenomenological implications.
In this article, we show how effective theories that parametrically violate the proposed conjectures can be UV-completed into theories that satisfy them.
The UV-completion is accessible through both the St\"uckelberg and Higgs mechanisms, with all dimensionless parameters taking $\mathcal{O}(1)$ values from the UV perspective.
These constructions feature an IR limit containing a light vector that is parametrically separated from any other massive states, and from any cut-off scale mandated by quantum gravity consistency requirements. Moreover, the cut-off--to--vector--mass ratio remains parametrically large even in the decoupling limit in which all other massive states (including any scalar excitations) become arbitrarily heavy.
We discuss how apparently strong constraints imposed by the proposed conjectures on phenomenologically interesting models, including specific production mechanisms of dark photon dark matter, are thereby circumvented.
}

\begin{document}

\maketitle

\section{Introduction}

The Swampland program aims to identify a well-defined set of conditions that effective field theories need to satisfy in order to be compatible with a UV-completion into a theory of quantum gravity \cite{Vafa:2005ui,ArkaniHamed:2006dz,Adams:2006sv,Ooguri:2006in,Brennan:2017rbf}.
Some of the best-known conjectures in the list of Swampland criteria are based on arguments regarding extremal black hole decay or absence of black hole remnants, such as e.g.~the weak gravity conjecture (WGC) \cite{ArkaniHamed:2006dz}, or the absence of global symmetries \cite{Kamionkowski:1992mf,Holman:1992us,Kallosh:1995hi,Banks:2010zn}. String theory then provides a playing field to look for counterexamples, gain intuition that allows to sharpen the various versions of these conjectures \cite{Heidenreich:2015nta,Heidenreich:2016aqi,Montero:2016tif,Andriolo:2018lvp,Heidenreich:2017sim}, or propose new ones.

In \cite{Reece}, two new additions to the list of Swampland criteria were proposed, aimed at effective theories containing $U(1)$ gauge bosons whose mass arises through the St\"uckelberg mechanism. Specifically, \cite{Reece} argues that in a theory featuring a vector with a St\"uckelberg mass $m_\gamma = g f_\theta$, with $g$ the corresponding gauge coupling, the following statements hold:
\begin{itemize}
	\item[(\emph{1})] \emph{`St\"uckelberg cut-off' conjecture}: there is a cut-off at the scale $\Lambda \lesssim \sqrt{f_\theta M_{Pl}}$, beyond which the effective field theory description breaks down.
	\item[(\emph{2})] \emph{`Radial mode' conjecture}: there is a dynamical scalar degree of freedom present at the scale $m_\sigma \lesssim f_\theta$.
\end{itemize}
These are adaptations of conjectures that already exist as applied to fundamental axion fields $\theta$ with period $2 \pi f_\theta$ \cite{Ooguri:2006in}.
The main insight of \cite{Reece} is to argue that they should also apply to massive vectors when the mass arises through the St\"uckelberg mechanism, since in such case the longitudinal mode of the vector may be regarded as a compact axion. 

Motivation behind conjecture (\emph{1}) rests on the observation that, in known string theory constructions, the point in moduli space at which the period of a compact axion vanishes -- and therefore the vector mass in the St\"uckelberg mechanism -- lies infinitely far away from any other point with non-zero period. This, together with conjecture no.~$2$ in \cite{Ooguri:2006in}, which states that the low energy effective theory defined around a given point in moduli space only remains consistent within a finite distance of such point, suggest that the limit $f_\theta \rightarrow 0$ should be non-smooth, and that the effective theory should have a cut-off scale that vanishes in that limit.
The specific form $\Lambda \lesssim \sqrt{f_\theta M_{Pl}}$ is obtained by applying the WGC to the theory dual to the scalar $\theta$, namely a $U(1)$ gauge theory with an associated 2-form gauge potential, and it is further motivated by arguments regarding the absence of black hole remnants \cite{Bowick:1988xh,Hebecker:2017uix}.

Regarding conjecture (\emph{2}), the idea that there must be a radial mode
accompanying the axion whose mass is not arbitrarily heavy
was first proposed in \cite{Ooguri:2006in} (conjecture no.~4).
Refinements of this conjecture in \cite{Reece}, through arguments based on the expectation that instanton effects will break the continuous shift symmetry of the axion field,
further justify the upper bound $m_\sigma \lesssim f_\theta$.
Although this may well be the situation \emph{in general} (such as e.g.~if the axion couples to the topological term of a non-abelian gauge theory, or in specific string constructions), this precise bound is more a statement about consistent interacting effective field theories than it is about a consistent theory of quantum gravity. In particular, conjecture (\emph{2}) is peculiar as a statement about quantum gravity insofar as, unlike conjecture (\emph{1}), taking the limit $M_{Pl} \rightarrow \infty$ does not decouple the radial mode.
Regardless of the precise form that conjecture (\emph{2}) should take {\it qua} a consistency condition of quantum gravity, we nevertheless entertain it in the following -- not least because we will also be considering Higgs theories, for which the conjecture obviously holds.

The conjectures in \cite{Reece} lead to (at least) two phenomenologically relevant consequences.
First, in conjunction with current experimental upper bounds, they imply that the Standard Model (SM) photon must be completely massless.
Second, these conjectures have the potential to significantly constrain the region of parameter space that is realizable for models of dark photon dark matter. Specifically, if the dark matter relic abundance of dark photons is produced through inflationary fluctuations, as proposed in \cite{Graham:2015rva}, the conjectures of \cite{Reece} imply that dark photon masses below $\sim 10 \ {\rm eV}$ may be inconsistent with a further UV-completion into a theory of quantum gravity.
If true, this would wipe out a significant region of the relevant parameter space in such theories, since from a low energy perspective dark photon masses as low as $\sim \mu {\rm eV}$ are consistent with a dark matter abundance generated from inflationary fluctuations.

In this work, we robustly demonstrate that the conjectures in \cite{Reece}, even if true in the UV, need not imply phenomenological consequences in the IR. That is, low energy effective field theories can appear to parametrically violate the conjectures, thereby removing phenomenological constraints. Our counter-example unsurprisingly takes the form of a small modification of the clockwork mechanism of \cite{Choi:2015fiu,Kaplan:2015fuy} as applied to vectors \cite{Saraswat:2016eaz}, and for reasons that will become apparent in the following we refer to it as `broken clockwork'.
Crucially, the broken clockwork constructions we present feature a low energy effective theory with the following properties:
\begin{itemize}
	\item[($i$)] it contains a massive photon that is parametrically below any other massive degrees of freedom, in particular any radial mode,
	\item[($ii$)] it allows for a decoupling limit in which scalar excitations (and other massive states) become arbitrarily heavy, but a parametric 		separation of 	scales remains between the photon mass and any cut-off scale mandated by quantum gravity arguments,
	\item[($iii$)] despite parametrically violating conjectures (\emph{1}) and (\emph{2}), it can be UV-completed into a theory that satisfies them, and the UV-completion 	can be implemented both through the Higgs and St\"uckelberg mechanisms, and
	\item[($iv$)] properties ($i$)-($iii$) hold for $\mathcal{O}(1)$ values of the UV parameters.
\end{itemize}

The rest of this paper is organized as follows.
In section \ref{sec:review} we review the clockwork mechanism as applied to vectors, as well as the properties of vector clockwork theories in the context of the WGC (following \cite{Saraswat:2016eaz}), and show how the vector clockwork construction can also be UV-completed through the St\"uckelberg mechanism.
In section \ref{sec:massivecw} we present how these constructions can be modified in order to obtain a theory that satisfies properties ($i$)-($iv$).
Section \ref{sec:pheno} revisits the phenomenological implications of conjectures (\emph{1}) and (\emph{2}), and discusses how broken clockwork models circumvent the constraints of \cite{Reece}.
We summarize our conclusions in section \ref{sec:conclusions}.

\section{Vector clockwork, and the WGC}
\label{sec:review}

We begin by reviewing the original clockwork mechanism of \cite{Choi:2015fiu,Kaplan:2015fuy} as applied to abelian gauge theories in section~\ref{sec:vectorCWhiggs}, as well as the status of the WGC in the context of these constructions, following \cite{Saraswat:2016eaz}.
In section~\ref{sec:vectorCWstuck} we present a St\"uckelberg UV-completion of vector clockwork.

\subsection{Vector clockwork through Higgsing}
\label{sec:vectorCWhiggs}

The discrete version of the clockwork construction as applied to vectors consists of a theory with a gauge group that is a product of $N+1$ independent $U(1)$ factors, $G = \prod_{j=0}^N U(1)_j$,
with a low energy effective lagrangian describing the gauge sector of the form \cite{Saraswat:2016eaz,Giudice:2016yja}
\begin{equation}
	\mathcal{L} = - \frac{1}{4} \sum_{j=0}^N F_{j \mu \nu}^2 + \frac{1}{2} \sum_{j=0}^{N-1} m^2 (A_{j \mu} - q A_{j+1 \mu})^2 \ ,
\label{eq:lagrangiancw0}
\end{equation}
where $q$ is a dimensionless quantity that needs to be $q \neq 1$ for clockwork to operate, and $m^2$ is a mass-squared parameter that breaks $N$ of the $N+1$ $U(1)$'s. Without loss of generality, here we will consider the case $q > 1$.
Compactness requires charge be quantized, and for simplicity we assume the charge quantum $g$ is the same for all $U(1)_j$ factors.

\begin{figure}[h]
    \centering
    \includegraphics[scale=0.65]{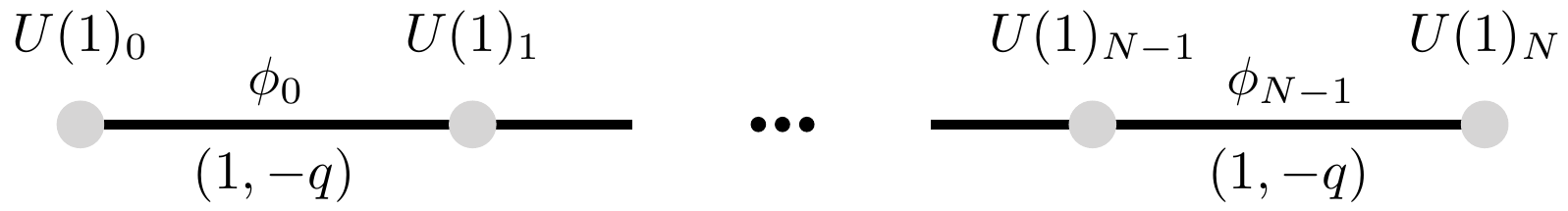}
\caption{Quiver diagram for the vector clockwork construction. Circles denote $U(1)$ gauge groups, and lines represent complex scalar fields charged under adjacent groups.}
\label{fig:quiver}
\end{figure}

A partial UV-completion to Eq.(\ref{eq:lagrangiancw0}) is achieved by introducing $N$ Higgs fields $\phi_j$ carrying charges $1$ and $-q$ under the $U(1)_j$ and $U(1)_{j+1}$ factors.
The corresponding lagrangian then reads
\begin{equation}
	\mathcal{L} = - \frac{1}{4} \sum_{j=0}^N F_{j \mu \nu}^2 + \sum_{j=0}^{N-1} \left( | D_\mu \phi_j |^2 - V (\phi_j) \right) \ ,
\label{eq:lagrangiancw}
\end{equation}
where $D_\mu \phi_j = \partial_\mu \phi_j - g (A_{j} - q A_{j + 1})_\mu \phi_j$, and $V (\phi_j)$ denotes a non-trivial potential for each complex scalar such that $\langle |\phi_j| \rangle = v / \sqrt{2}$.
Eq.(\ref{eq:lagrangiancw0}) with $m^2 = g^2 v^2$ then corresponds to the effective lagrangian describing the gauge sector of the theory, in unitary gauge.
(It is useful to think of this construction in terms of a quiver theory, as depicted in Figure~\ref{fig:quiver}.)

Upon diagonalization, the spectrum of mass eigenstates contains a massless vector, corresponding to the $U(1)$ factor that remains unbroken, which we refer to as $U(1)_{cw}$.
In terms of the gauge fields associated to the individual lattice sites, it is given by
\begin{equation}
	A_\gamma = \frac{1}{\mathcal N} \sum_{j=0}^N q^{N-j} A_j \sim \sum_{j=0}^N \frac{1}{q^j} A_j\ ,
\end{equation}
where $\mathcal N = \sqrt{q^{2N} + \cdots + q^2 + 1} \sim q^N$.
The massless mode $A_\gamma$ is a linear combination involving the $N+1$ gauge bosons of all the original $U(1)$ factors, but with exponentially distributed coefficients.
In terms of the quiver theory of Figure~\ref{fig:quiver}, it is exponentially localized towards the $j=0$ site, and has exponentially suppressed overlap with states localized on the site $j=N$.

\begin{figure}[h]
    \centering
    \includegraphics[scale=0.65]{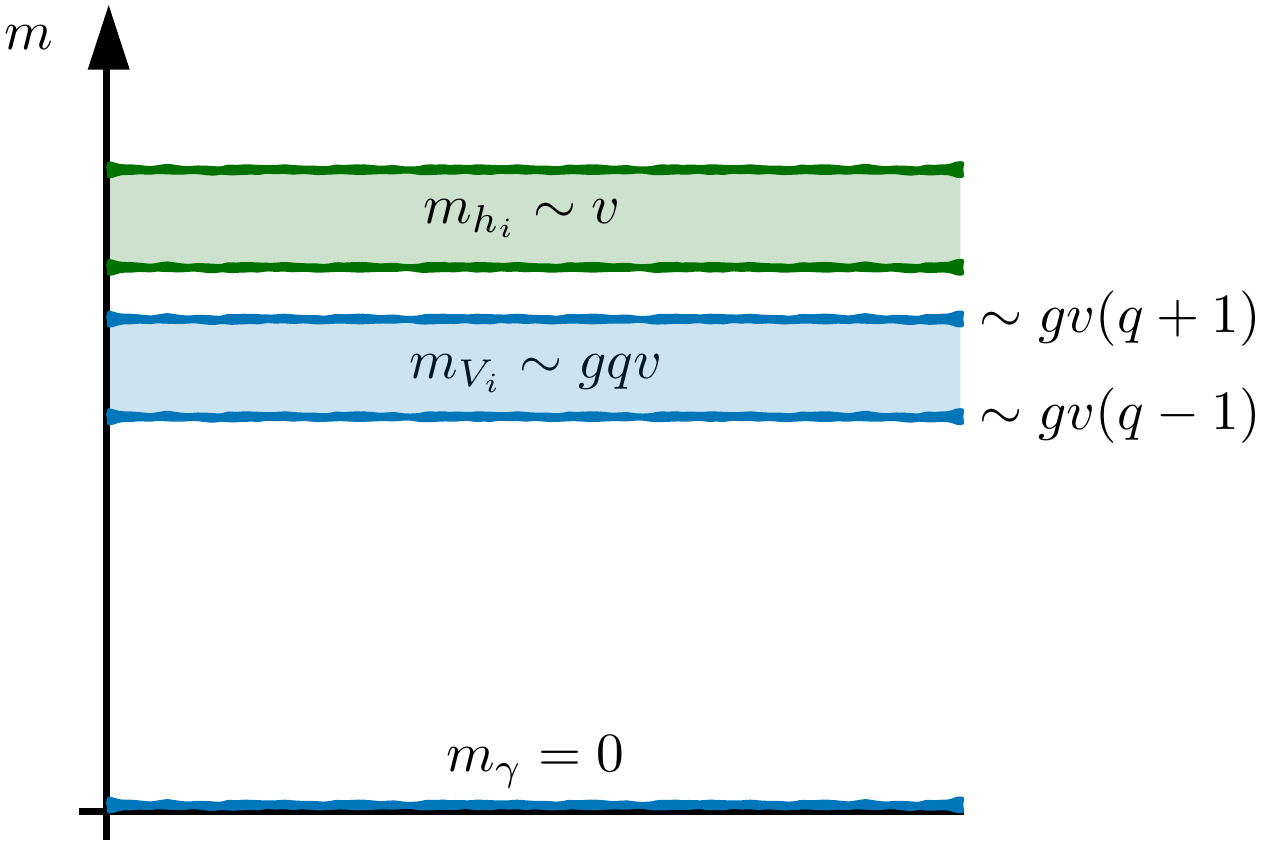}
\caption{Typical mass spectrum of vector clockwork constructions. A massless photon remains in the spectrum, followed by $N$ massive vectors at scale $m_{V_i} \sim g q v \lesssim v$.
In the Higgs UV-completion of the model, $N$ scalar fields are also present, with masses $m_{h_i} \sim v$.}
\label{fig:spectrumcw}
\end{figure}

As discussed in \cite{Giudice:2016yja}, the $N$ massive vectors resulting from this symmetry-breaking pattern are also linear combinations of all the fields in the quiver, but do not exhibit strong localization.
Their masses are at the scale $m_{V_i} \sim g q v$, with the lightest massive mode appearing at $m_{V_1} \approx (q-1) g v$, and the heaviest at $m_{V_N} \approx (q+1) g v$.
This mass spectrum is characteristic of clockwork theories: a massless mode followed by a band of $\mathcal{O} (gv)$ where the $N$ massive modes lie.
In this particular UV-completion, $N$ real scalar fields, corresponding to the radial modes of the $\phi_j$ fields, will also be part of the spectrum, with masses of order $\sim v$ (up to quartic couplings).
This is schematically depicted in Figure~\ref{fig:spectrumcw}.
Given the perturbativity constraint $g q \lesssim 1$, the massive vector modes could well lie below the scale $v$.
However, we will assume throughout that $gq \sim 1$, and use $m_\sigma \sim v$ as a proxy for the scale at which massive states, both scalars and vectors, are present.

Further, as a result of charge being quantized, the symmetry breaking pattern of the theory is not quite $U(1)^{N+1} \rightarrow U(1)_{cw}$, but rather
\begin{equation}
	U(1)^{N+1} \rightarrow U(1)_{cw} \times \left( \mathbb{Z}_{p_1} \times \cdots \times \mathbb{Z}_{p_N} \right) \ ,
\end{equation}
i.e.~$N$ discrete gauge symmetries remain unbroken.\footnote{For general $N$, $p_1 = p_N = q^{2 N} + \cdots + q^2 + 1$, whereas the order of the other $N-2$ discrete groups will typically be smaller, their exact value depending on the specific choice of $q$ and $N$.}
The leftover discrete symmetries are expressed in the spectrum of the theory through the presence of solitonic degrees of freedom:
flux tubes, or cosmic strings, inside of which magnetic flux of the broken gauge directions remains confined \cite{Nielsen:1973cs}.
There will be $N$ different types of strings, all with tension $T \sim v^2$.
Notice that the presence and properties of these strings can be understood purely from a semi-classical analysis of Eq.(\ref{eq:lagrangiancw}) (see e.g.~\cite{Preskill:1986kp}),
and no new degrees of freedom, or any other modification of the theory, need to be introduced at scale $\sqrt{T} \sim v$.

Many of the interesting properties of clockwork theories stem from the exponential localization of its 0-mode. In particular, a matter field $\eta_N$ carrying unit charge $g$ under the gauge group $U(1)_N$
will couple to the massless mode with strength
\begin{equation}
	g_\gamma = \frac{g}{\mathcal{N}} \sim \frac{g}{q^N} \ll g \ ,
\label{eq:geff}
\end{equation}
and defines the charge quantum of the unbroken $U(1)$ factor.
In general, a state $\eta_j$ carrying unit charge under the $j$-th gauge group, carries charge $q^j$ (in units of $g_\gamma$) under $U(1)_{cw}$.
Compactness of the theory in the broken phase is therefore guaranteed by assuming the original $N+1$ $U(1)$ factors are also compact.

As first noted in \cite{Saraswat:2016eaz}, the exponential localization of the 0-mode across the quiver, and specifically Eq.(\ref{eq:geff}) as a direct consequence,
endows this theory with rather unusual properties in the context of the WGC.
In particular, specific versions of the conjecture satisfied by the UV theory (that is, before Higgsing), may be parametrically violated in the IR.
For instance, the `unit-charge' version of the conjecture may be satisfied in the UV by demanding all states $\eta_j$ appear at a scale $m_\eta$ satisfying
\footnote{For simplicity we assume all states $\eta_j$ have similar mass $\sim m_\eta$.}
\footnote{The right-hand-side of Eq.(\ref{eq:WGCUV}) should include an extra factor of $1/\sqrt{4 \pi}$, as well as $1/\sqrt{N+1}$ from considering a theory involving several $U(1)$ factors \cite{Cheung:2014vva}. However, since we will be focused in cases with $N = \mathcal{O}(1)$, these factors are irrelevant for our discussion, and we neglect them in the following.}
\begin{equation}
	m_\eta \lesssim \Lambda \sim g M_{Pl} \ ,
\label{eq:WGCUV}
\end{equation}
where $g$ is subject to the perturbativity constraint $g q \lesssim 1$.
For $\mathcal{O}(1)$ values of $q$, this is not a strong constraint, and if $g$ is not very small, $\Lambda$ may not be far below $M_{Pl}$.
Thus, the cut-off scale defined by the WGC, below which electrically charged states must be present in the theory, can easily be above all the massive vector and scalar excitations depicted in Figure~\ref{fig:spectrumcw},
and therefore well out of reach of any low energy effective description.
\footnote{Sublattice and Tower versions of the WGC further suggest that local effective field theory {\it completely} breaks down at a scale of order $g^{1/3} M_{Pl}$ \cite{Heidenreich:2016aqi, Heidenreich:2017sim}. However, as noted in \cite{Heidenreich:2017sim}, these considerations only strictly apply in dimensions greater than four, and while some version of these conjectures may persist in purely four-dimensional theories, its precise form remains unclear. Since our conclusions are independent of the specific form of the cut-off, we will stick with the unit-charge or magnetic form, $\Lambda \sim g M_{Pl}$, unless otherwise noted.}

However, in light of Eq.(\ref{eq:geff}) it is clear that imposing this version of the conjecture in the UV does not imply that the same version is satisfied by the IR theory.
In the broken phase, a state with mass $m_\eta$ and $U(1)_{cw}$ charge $Q$ (in units of $g_\gamma$) is super-extremal if
\begin{equation}
	m_\eta \lesssim Q g_\gamma M_{Pl} \sim \frac{Q}{q^N} g M_{Pl} \ .
\label{eq:WGCIR}
\end{equation}
Eq.(\ref{eq:WGCUV}) obviously implies Eq.(\ref{eq:WGCIR}) for $Q=q^N$, and therefore the state $\eta_0$, carrying unit charge under $U(1)_0$, is guaranteed to satisfy the super-extremality condition in the IR theory.
However, if $m_\eta$ is not too far below $\Lambda$, this may indeed be the only super-extremal state present in the Higgsed phase.
As a result, from a UV theory that satisfies the stringent unit-charge version of the conjecture, we recover an IR theory in which the only super-extremal state carries charge $q^N \gg 1$,
parametrically violating the version of the conjecture imposed in the unbroken phase.
\footnote{As discussed in \cite{Saraswat:2016eaz}, this state of affairs is however not a problem as far as arguments regarding the decay of extremal black holes are concerned: even though black holes
can only lose charge modulo $q^N$, there are no controlled extremal black hole solutions (that is, with mass above $M_{Pl}$) carrying smaller charge.}

The discussion of the magnetic version of the conjecture proceeds along similar lines \cite{Saraswat:2016eaz}.
Assuming the magnetic WGC holds in the unbroken phase, $\Lambda \sim g M_{Pl}$ corresponds to the scale at which the UV theory needs to be modified in order to account for the presence of magnetic monopoles charged under the $U(1)_j$ factors \cite{ArkaniHamed:2006dz}.
Applying the same version of the conjecture to the IR theory would suggest that new physics should therefore be present at a scale $g_\gamma M_{Pl} \sim \Lambda / q^N \ll \Lambda$
to account for monopoles carrying $U(1)_{cw}$ magnetic charge.
But no modification of the theory at scale $g_\gamma M_{Pl}$ is required.
Instead, IR monopole configurations may be built out of individual monopoles carrying magnetic charge of the $U(1)_j$ factors, connected through flux tubes.
In this way, a finite energy configuration carrying $U(1)_{cw}$ magnetic charge can be built, and as shown in \cite{Saraswat:2016eaz} the unit-charge monopole (that is, with charge $2 \pi / g_\gamma$) is not a black hole.
Thus, the IR theory parametrically violates the magnetic form of the WGC, although it is satisfied in the UV.

\subsection{Vector clockwork through the St\"uckelberg mechanism}
\label{sec:vectorCWstuck}

The low energy effective lagrangian of Eq.(\ref{eq:lagrangiancw0}) can also be UV-completed through the St\"uckelberg mechanism, by introducing $N$ axion fields $\theta_j$, each transforming non-trivially
under the gauge transformation corresponding to the linear combination $A_j - q A_{j+1}$.

A compact axion can be dualized into a 2-form gauge field associated with an abelian gauge group.
For the case at hand, the St\"uckelberg UV-completion of the vector clockwork construction requires introducing $N$ such 2-forms, $B_{j}$, and the corresponding lagrangian can be written as
\begin{equation}
	\mathcal{L} = - \frac{1}{4} \sum_{j=0}^N F_{j \mu \nu}^2 + \sum_{j=0}^{N-1} \left( \frac{1}{12} H_{j \mu \nu \sigma}^2
		+ \frac{g f_\theta}{4} \varepsilon^{\mu \nu \rho \sigma} (F_{j \mu \nu} - q F_{j+1 \mu \nu} ) B_{j \rho \sigma} \right) \ ,
\label{eq:lagCWstuck}
\end{equation}
where $H_j = d B_j$ are the corresponding 3-form field strengths, and are related to the scalars $\theta_j$ through the Hodge dual operation $H_j = * d \theta_j$.
The 2-forms $B_j$ couple to fundamental string currents in units of $2 \pi f_\theta$, which coincides with the periodicity of the scalars $\theta_j$. As the $A_j$ and $B_j$ each are associated with compact abelian gauge groups, the parameter $q$ is necessarily quantized in $\mathbb{Z}$. The clockwork form of the $B_j \wedge (F_j - q F_{j+1})$ couplings can be understood simply as the dual of the $(1,-q)$ ``charges'' dictating shifts of the axion fields $\theta_j$ under gauge transformations of $(A_j, A_{j+1})$.

The $B_j \wedge (F_j - q F_{j+1})$ couplings of Eq.(\ref{eq:lagCWstuck}) can be rewritten in terms of the $H_j$, which in turn may be eliminated through the corresponding equations of motion:
\begin{equation}
	H_j^{\mu \rho \sigma} = g f_\theta \varepsilon^{\mu \rho \sigma \nu} (A_j - q A_{j+1})_\nu \ .
\label{eq:Heom}
\end{equation}
Plugging Eq.(\ref{eq:Heom}) back into Eq.(\ref{eq:lagCWstuck}) we precisely recover Eq.(\ref{eq:lagrangiancw0}) with $m^2 = g^2 f_\theta^2$.

Na\"ively, the mass spectrum of this UV-completion is as in Figure~\ref{fig:spectrumcw} except that there are no massive scalars.
However, if we accept the `radial mode' conjecture of \cite{Reece}, every $\theta_j$ must be accompanied by a scalar excitation appearing at a scale $m_\sigma$ not far above $f_\theta$.
In such case, the spectrum of the theory is then in fact identical to that of Figure~\ref{fig:spectrumcw}, after the obvious replacement $v \rightarrow f_\theta$.

A further consideration stems from applying the WGC to the abelian gauge groups associated to the 2-forms.
Using the version of the conjecture appropriate for higher form fields \cite{ArkaniHamed:2006dz}, one concludes that the tension $T$ of fundamental strings coupling to the $B_j$'s must satisfy 
\footnote{Eq.(\ref{eq:tension}) would correspond to the unit-charge version of the conjecture as applied to 2-form abelian gauge theories, up to $\mathcal{O}(1)$ factors we are not keeping track of.}
\begin{equation}
	T \lesssim f_\theta M_{Pl}  \ .
\label{eq:tension}
\end{equation}

Much like the flux tubes present in Higgs theories, these strings will carry magnetic flux of the broken gauge directions.
This can be seen by introducing string currents $\Sigma_j$ with couplings of the form $\mathcal{L} \propto 2 \pi f_\theta B_{j \mu \nu} \Sigma_j^{\mu \nu}$.
Taking the divergence of the equations of motion obtained by varying with respect to the $B_j$'s, one indeed finds
\begin{equation}
	\partial_\mu ({\tilde F}_j - q {\tilde F}_{j+1})^{\mu \nu} \propto \frac{2 \pi}{g} \partial_\mu \Sigma_j^{\mu \nu} \ ,
\end{equation}
where ${\tilde F}_j \equiv * F_j$.
However, unlike Higgs cosmic strings which can be understood semi-classically, with no need for extra degrees of freedom beyond those featured in Eq.(\ref{eq:lagrangiancw}), the tension of St\"uckelberg strings defines a scale at which the field content of the theory needs to be extended. In this sense, $\sqrt{T} \sim \sqrt{f_\theta M_{Pl}}$ may be regarded as a cut-off scale, as mandated by the WGC \cite{Hebecker:2017uix}.

The discussion of the WGC as applied to the 1-form gauge sector is identical to the situation in the Higgs UV-completion, with fundamental strings now playing the role of flux tubes in building the IR monopoles.
Depending on the sizes of $g$ and $f_\theta$, the WGC cut-off as applied to the St\"uckelberg UV-completion will then be either
\begin{equation}
	\Lambda \sim g M_{Pl} \quad {\rm or} \quad \sqrt{f_\theta M_{Pl}} \ .
\end{equation}

\section{Broken vector clockwork}
\label{sec:massivecw}

\subsection{Parametrically light vectors from a broken clockwork}
\label{sec:brokencw}

We consider a modification of the original clockwork construction by adding to the lagrangian of Eq.(\ref{eq:lagrangiancw0}) an extra term of the form:
\begin{equation}
	\mathcal{L} \supset \frac{1}{2} m^2 A_N^2 \ ,
\label{eq:Lcwbreaking}
\end{equation}
which breaks the $U(1)_{cw}$ symmetry.
This term can be implemented through the addition of an extra Higgs field carrying unit charge under the last gauge group of the quiver, i.e.~
\begin{equation}
	\mathcal{L} \supset |D_\mu \phi_N|^2 - V(|\phi_N|) \ ,
\end{equation}
with $D_\mu \phi_N = \partial_\mu \phi_N - i g A_{N \mu} \phi_N$, and $V(|\phi_N|)$ a non-trivial potential such that $\langle |\phi_N| \rangle = v / \sqrt{2}$.
Or by introducing an additional 2-form field in the St\"uckelberg case, with couplings:
\begin{equation}
	\mathcal{L} \supset \frac{1}{12} H_{N \mu \nu \sigma}^2 + \frac{g f_\theta}{4} \varepsilon^{\mu \nu \rho \sigma} F_{N \mu \nu} B_{N \rho \sigma} \ .
\end{equation}
A symmetry breaking term as in Eq.(\ref{eq:Lcwbreaking}) arises from either of these choices, with $m^2 = g^2 v^2$ or $g^2 f_\theta^2$, for the Higgs and St\"uckelberg UV-completions respectively.
\footnote{Choosing the vev of the additional Higgs field $\phi_N$ (the periodicity of the additional St\"uckelberg axion $\theta_N$) to be different from  $v$ ($f_\theta$) by an
$\mathcal{O}(1)$ amount makes no qualitative difference to our conclusions.}
In the following, we will use notation appropriate to the Higgs UV-completion of the model, but all our results apply also in the St\"uckelberg case unless otherwise noted.

For a $2$-site model ($N=1$), the eigenvalues of the modified vector mass-squared matrix can straightforwardly be obtained analytically, and are given by
$m^2_{\gamma, V} = (g v)^2 \lambda_{\gamma,V}$, with
\begin{align}
	\lambda_\gamma & = \frac{1}{2} \left( q^2 + 2 - \sqrt{(q^2 + 2)^2 - 4} \right) = \frac{1}{q^2}  \left( 1 + \mathcal {O} \left( q^{-2} \right) \right) \sim \frac{1}{q^2} \ ,\\
	\lambda_V & = \frac{1}{2} \left( q^2 + 2 + \sqrt{(q^2 + 2)^2 - 4} \right) = q^2  \left( 1 + \mathcal {O} \left( q^{-2} \right) \right) \sim q^2 \ .
\end{align}
Thus, the masses of the two vectors are, parametrically:
\begin{equation}
	m_\gamma \equiv g v \sqrt{\lambda_\gamma} \sim \frac{g v}{q} \ , \qquad {\rm and} \qquad m_V \equiv g v \sqrt{\lambda_V} \sim g q v \ .
\end{equation}

The theory no longer contains a massless mode, but
otherwise the spectrum of states is very much like in the standard clockwork construction, with a second massive vector at the scale $m_V \sim g q v \lesssim v$, and scalar excitations appearing at the scale $\sim v$.
The separation of scales between the light vector and heavier states with mass $m_\sigma \sim v$ (or $f_\theta$), is given by
\begin{equation}
	\frac{m_\sigma}{m_\gamma} \sim \frac{q}{g} \gtrsim q^2 \ ,
\end{equation}
where in the last step we have used the perturbativity requirement $g q \lesssim 1$.
However, this separation of scales can only be made parametrically large if one chooses $q \gg 1$, which may seem \emph{ad hoc}, and could potentially frustrate the theory's embedding into a full UV-completion \cite{Heidenreich:2015wga,Ibanez:2017vfl}.

Interestingly, the separation of scales increases exponentially by increasing the number of sites.
The mass of the lightest vector in the general case containing $N+1$ sites is
\begin{equation}
	m_\gamma \sim \frac{g v}{q^N} \ ,
\label{eq:mgamma}
\end{equation}
whereas the heavier vector spectrum is left almost unchanged.

While complete analytic expressions for the eigenvalues and eigenvectors are unilluminating, it is straightforward to see the effect of the mass deformation in Eq.(\ref{eq:Lcwbreaking}) on the $(N+1)$-site clockwork spectrum at first order in 
$m^2/(g q v)^2 \sim 1 / q^2$.
The mass-squared eigenvalues to first order in this perturbation are
\begin{equation}
	m_{V_i}^2 = \hat m_{V_i}^2 + m^2 O_{N i}^2
\end{equation}
where $\hat m_{V_i}$ denote the unbroken clockwork mass eigenvalues (as in section~\ref{sec:vectorCWhiggs}), and \cite{Giudice:2016yja}
\begin{eqnarray}
O_{i0} = \frac{q^{N-i}}{\mathcal{N}} \ , \hspace{1cm} O_{ij} = \mathcal{N}_j \left[q \sin \frac{ij \pi}{N+1}  -\sin \frac{(i+1) j \pi}{N+1} \right] \\
\mathcal{N} = \sqrt{q^{2N} + \cdots + q^2 + 1} \ , \hspace{1cm} \mathcal{N}_j = \sqrt{\frac{2 g^2 v^2}{(N+1) \hat m_{V_j}^2}}
\end{eqnarray}
is the orthogonal matrix relating the gauge and mass eigenbases for the unperturbed clockwork, $\hat A = O \hat V$. In particular, the mass of the lightest eigenstate at this order is precisely that of Eq.(\ref{eq:mgamma}), while the perturbations to the heavier mass eigenstates are suppressed by $1/N$. The lightest mass eigenstate remains exponentially localized; to first order in
$m^2/(g q v)^2 \sim 1 / q^2$
it is related to the unperturbed clockwork eigenvectors via
\begin{equation}
A_\gamma \propto \hat A_\gamma - \sum_i \frac{m^2}{\hat m_{V_i}^2} O_{Ni} O_{N0} \hat V_i \ .
\end{equation} 
In particular, the admixture of the lightest mode with the heavier unperturbed mass eigenstates is suppressed by a factor of $O_{N0} \sim 1/q^N$, preserving the localization of the zero mode observed in the unperturbed theory.

\begin{figure}[h]
    \centering
    \includegraphics[scale=0.65]{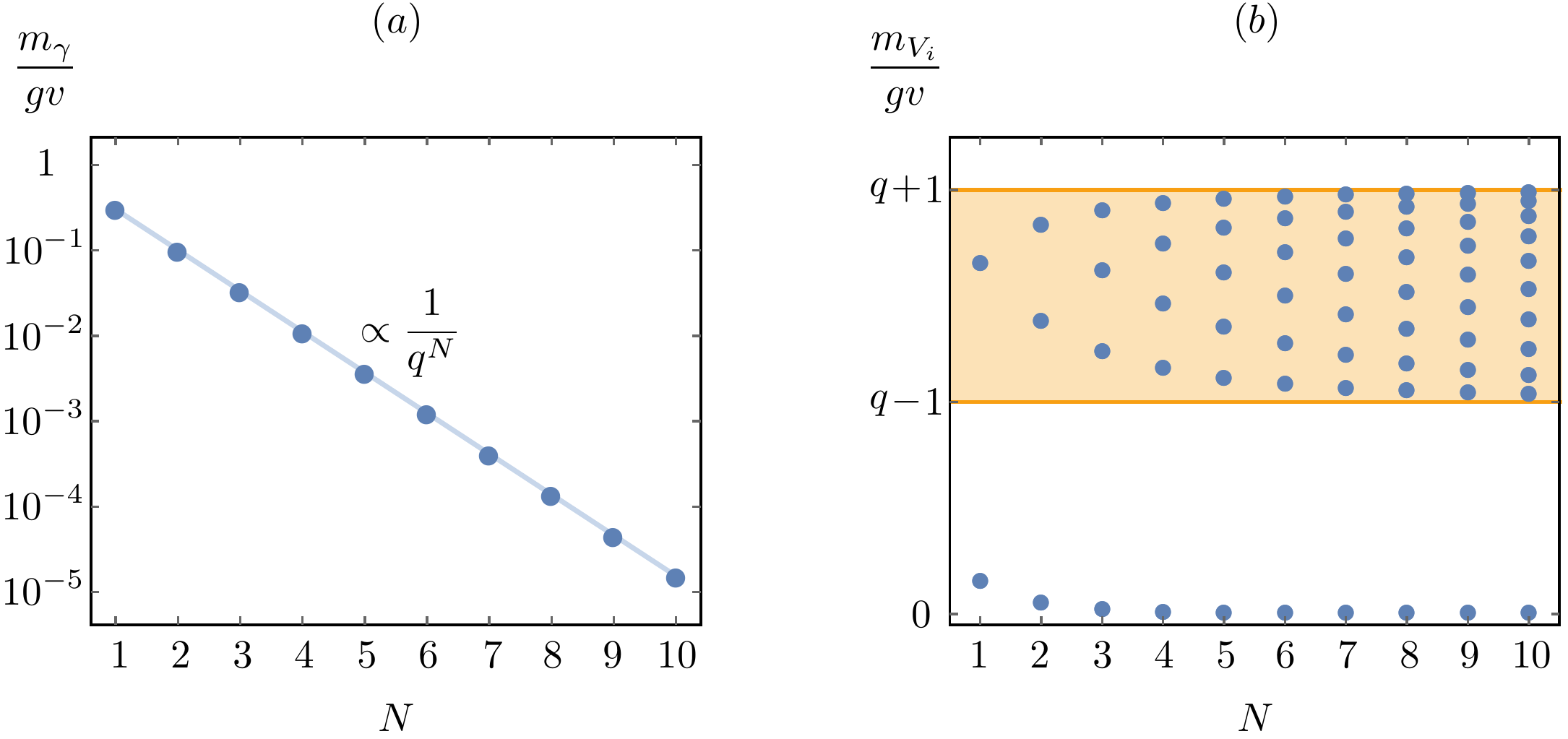}
\caption{Typical mass spectrum of the broken clockwork constructions discussed here. $(a)$ Mass of the lightest vector as a function of $N$.
		$(b)$ Masses of the $N$ heavier vector modes. As in the standard (unbroken) clockwork model, their masses range from approximately $(q-1)gv$ to $(q+1)gv$ (orange band).
		In both figures, $q=3$ for illustration.}
\label{fig:masses}
\end{figure}

\begin{figure}[h]
    \centering
    \includegraphics[scale=0.65]{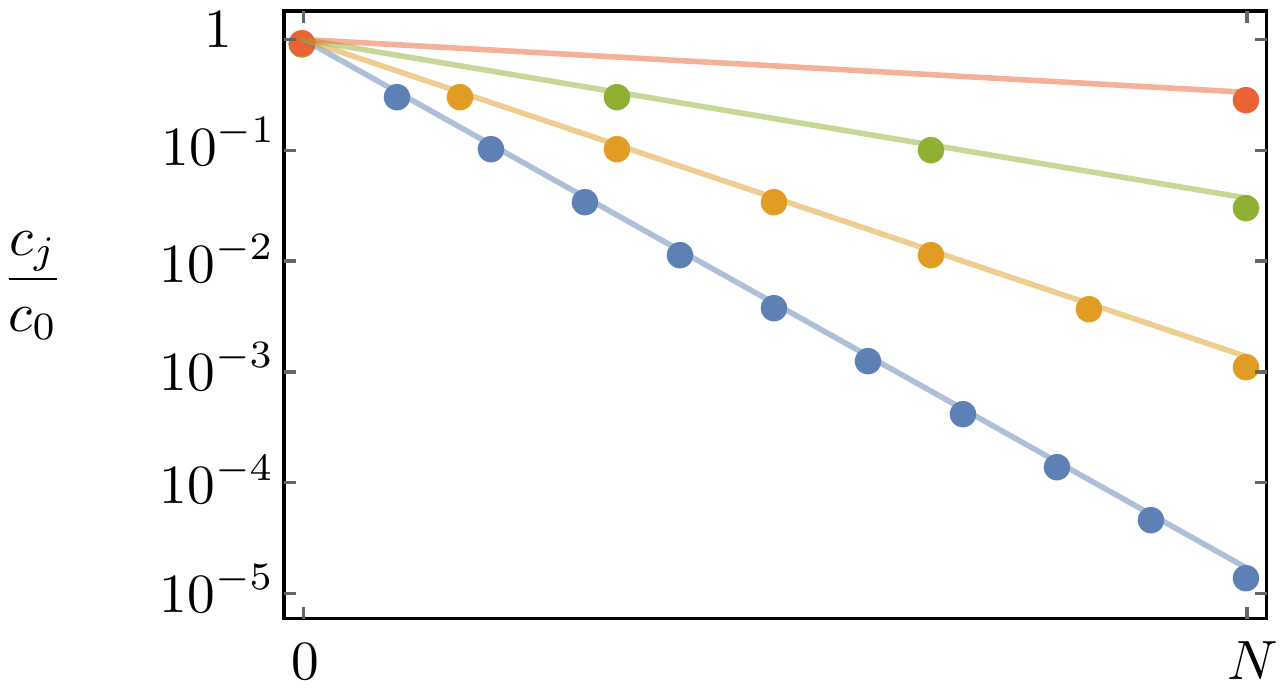}
\caption{Profile of lightest vector for $N=1$ (red), 3 (green), 6 (orange), and 10 (blue), with $q=3$ for illustration,
	with $A_\gamma = \sum_{j=0}^N c_j A_j$.
	Continuous lines correspond to the exponential localization of the standard clockwork construction where $c_j / c_0 = 1/q^{j}$.}
\label{fig:profilegamma}
\end{figure}

The stated parametric behaviour apparent from perturbation theory can be seen more robustly in Figure~\ref{fig:masses}, which shows the exact numerical spectrum as a function of the number of sites.
The eigenvector spectrum also very closely matches that of the unbroken clockwork construction. In particular, the lightest vector exhibits the same strong localization towards one of the sites in the quiver, as expected; this is illustrated in Figure~\ref{fig:profilegamma}.

The scaling behaviour in Eq.(\ref{eq:mgamma}) allows for a separation of scales between the lightest vector, and all the other massive modes appearing at scale $m_\sigma \sim v$ (or $f_\theta$):
\begin{equation}
	\frac{m_\sigma}{m_\gamma} \sim \frac{q^N}{g} \gtrsim q^{N+1} \ ,
\label{eq:sigmagammaratio}
\end{equation}
which may be parametrically large even for $\mathcal{O}(1)$ values of the UV parameters $q$ and $N$.

\subsection{Status of the WGC}

In section~\ref{sec:brokencw} we have discussed how a broken clockwork construction can feature a large separation of scales between the lightest state and any other massive excitations, even for $\mathcal{O}(1)$ values of the underlying UV parameters, as illustrated through Eq.(\ref{eq:sigmagammaratio}).
However, if we are willing to consider very small parameter values, so long as they are technically natural, we could have just as well considered a theory with a single $U(1)$ and a Higgs field (or St\"uckelberg axion) carrying the unit of charge.
In this case, we would find
\begin{equation}
	\frac{m_\sigma}{m_\gamma} \sim \frac{\sqrt{\lambda}}{g} \gg 1 \qquad {\rm for} \qquad g \ll 1 \ ,
\end{equation}
with $\lambda$ a quartic coupling controlling the radial mode mass in the Higgs or St\"uckelberg constructions, satisfying the perturbativity requirement $\lambda \lesssim 4 \pi$.
In fact, what we have done in our broken clockwork construction is first use the standard clockwork mechanism to generate a very small effective gauge coupling,
Eq.(\ref{eq:geff}), and then break it through a Higgs or St\"uckelberg axion field that carries the unit of charge under $U(1)_{cw}$.
Surely then, the conclusions regarding separation of scales in the two theories cannot be that different, and it is therefore a matter of taste whether one prefers a theory with very small couplings, or one with $\mathcal{O}(1)$ parameters but with an extended field content?
It turns out that the two theories are crucially different when considered in the context of the WGC, as we now discuss.

Some of the unusual features of vector clockwork theories with regards to the WGC are expressed also in the broken version of these constructions.
Specifically, a parametric separation of scales between the photon mass $m_\gamma$, and any cut-off scale required by WGC-like arguments remains, even in the decoupling limit in which all other massive states are taken arbitrarily heavy.
For instance, in either the Higgs or St\"uckelberg versions of the theory with $\Lambda \sim g M_{Pl}$, we have
\begin{equation}
	\frac{\Lambda}{m_\gamma} \lesssim \frac{g M_{Pl}}{ g v / q^N} = q^N \frac{M_{Pl}}{v} \xrightarrow{v \rightarrow M_{Pl}} q^N \gg 1 \ ,
\end{equation}
where $v$ should be replaced by $f_\theta$ in the St\"uckelberg case.

This result is largely independent of the specific form of the cut-off imposed in the UV theory. For example, if the parameters of the St\"uckelberg UV-completion are such that $\sqrt{f_\theta M_{Pl}}$ is below $g M_{Pl}$, we then find
\begin{equation}
	\frac{\Lambda}{m_\gamma} \lesssim \frac{\sqrt{f_\theta M_{Pl}}}{ g f_\theta / q^N} = \frac{q^N}{g} \sqrt{ \frac{M_{Pl}}{f_\theta} } \gtrsim \xrightarrow{f_\theta \rightarrow M_{Pl}} \frac{q^N}{g} \gtrsim q^{N+1} \gg 1 \ .
\end{equation}
Different values of $\Lambda$ arising from different versions of the conjecture as applied to the UV theory lead to different parametric dependence of the ratio $\Lambda / m_\gamma$ on the model parameters, but in any case a parametrically large ratio will remain even in the decoupling limit.

This is crucially different from the situation that arises from the spontaneous breaking of a single $U(1)$ gauge group through the vev of a Higgs carrying charge $g \ll 1$.
In this case, the WGC cut-off--to--photon--mass ratio behaves, parametrically
\begin{equation}
	\frac{\Lambda}{m_\gamma} \sim \frac{g M_{Pl}}{g v} = \frac{M_{Pl}}{v} \xrightarrow{v \rightarrow M_{Pl}} 1 \ ,
\end{equation}
i.e.~imposing the WGC in the unbroken phase precludes a decoupling limit.
This also holds for a St\"uckelberg construction, and since we are interested in $g$ small and $f_\theta$ large, the above expression trivially applies in this case after the substitution $v \rightarrow f_\theta$.

\section{Phenomenological implications}
\label{sec:pheno}

\subsection{Dark photon dark matter}

Massive vectors with small couplings to the SM degrees of freedom are ubiquitous in Beyond-the-Standard-Model physics.
They are a common occurrence in models of dark matter featuring extended dark sectors, and appear naturally in the context of string compactifications \cite{Arvanitaki:2009hb}. Moreover, if sufficiently stable, they can be the dark matter \cite{Nelson:2011sf} (see also \cite{Arias:2012az}).

In \cite{Graham:2015rva}, an attractive mechanism for generating dark photon dark matter was proposed, in which the correct relic abundance is produced through inflationary fluctuations. The mechanism is minimal, its only necessary ingredients being a massive vector and a period of inflation. Moreover, in order for the dark photon to account for all of the dark matter, its mass must be related to the scale of inflation as follows:
\begin{equation}
 	m_\gamma = 6 \ \mu {\rm eV} \left( \frac{10^{14} \ {\rm GeV}}{H_I} \right)^4 \ ,
\label{eq:mvsH}
\end{equation}
effectively making the mechanism a single parameter model.

For the mechanism to be successful, there must be no scalar fields with masses below $H_I$ (the Hubble scale during inflation) -- if there were, the isocurvature perturbations produced would be incompatible with CMB observations, and the model ruled out \cite{Fox:2004kb}.
Within an effective field theory framework, we would therefore expect the mechanism to be valid only if the mass of the vector is of St\"uckelberg type, or if the scalar excitation was parametrically decoupled in the Higgs case.
The minimality of the mechanism therefore appears as appealing as it is necessary.

In light of the conjectures in \cite{Reece}, however, this minimality may in fact be a curse rather than a blessing. In particular, the conjectured bounds in \cite{Reece} imply that even if the dark photon mass is of a St\"uckelberg type, dark photon masses below $\sim 10 \ {\rm eV}$ are challenging to realize in a theory of quantum gravity.
If taken at face value, this would significantly compromise a large region of parameter space in which the mechanism presented in \cite{Graham:2015rva} is applicable, and that will be explored by future experimental proposals \cite{Chaudhuri:2014dla} (see e.g.~Figure 6 in \cite{Graham:2015rva}). 
Thankfully, the apparent exponential violation of the conjectures in \cite{Reece} by effective field theories arising from broken clockwork reconciles this parameter space with UV-completion in a theory of quantum gravity. We now briefly review the logic presented in \cite{Reece}, before showing how the result is modified in the broken clockwork constructions of section~\ref{sec:brokencw}.

Taking the radial mode conjecture of \cite{Reece} as a reference, and demanding that no scalar excitations be present below the inflationary scale requires
\begin{equation}
	H_I \lesssim m_\sigma \ ,
\label{eq:radialmodeH}
\end{equation}
where $m_\sigma \lesssim v$ or $f_\theta$, for the Higgs and St\"uckelberg scenarios respectively.
Since we are interested in the regime of small dark photon mass, where $m_\gamma \ll H_I$, Eq.(\ref{eq:radialmodeH}) therefore requires that we take $g \ll 1$.
Further, for the calculation in \cite{Graham:2015rva} to be valid, $H_I$ must be below any cut-off scale beyond which the low energy effective theory breaks down.
In the regime of small gauge coupling, \cite{Reece} considers the Tower WGC cut-off $\Lambda \sim g^{1/3} M_{Pl}$.
In this case:
\begin{equation}
	H_I \lesssim \Lambda \lesssim g^{1/3} M_{Pl} \sim \left( \frac{m_\gamma}{H_I} \right)^{1/3} M_{Pl} \ ,
\label{eq:Hwgc}
\end{equation}
where in the last step we have used $g \sim m_\gamma / H_I$, saturating Eq.(\ref{eq:radialmodeH}) (this is the best case scenario, with the largest possible $H_I$ and so the smallest possible $m_\gamma$).
Plugging Eq.(\ref{eq:Hwgc}) back into Eq.(\ref{eq:mvsH}) translates into the constraint, parametrically:
\begin{equation}
	m_{\gamma} \gtrsim 10 \ {\rm eV} \ .
\end{equation}

We can repeat this exercise for the broken clockwork construction. Taking into account the relationship between the mass of the lightest vector and heavier states, as given through Eq.(\ref{eq:sigmagammaratio}), and assuming Eq.(\ref{eq:radialmodeH}) holds, we then have $g \sim q^N m_\gamma / H_I$.
This modifies Eq.(\ref{eq:Hwgc}) as 
\begin{equation}
	H_I \lesssim  \left( \frac{q^N m_\gamma}{H_I} \right)^{1/3} M_{Pl} \ ,
\end{equation}
and in turn weakens the lower bound on the dark photon mass:
\begin{equation}
	m_\gamma \gtrsim \frac{10 \ {\rm eV }}{q^{N/2}} \ ,
\end{equation}
which is $\sim \mu {\rm eV}$ for, for example, $q=3$ and $N \approx 16$.

Hence, the broken clockwork constructions discussed here allow for dark photon masses in the entire regime in which the dark photon can be the dark matter, all the way down to $m_\gamma \sim \mu {\rm eV}$.
Moreover, since the discussion above focused on the radial mode conjecture of \cite{Reece}, our conclusions are valid both in the St\"uckelberg and Higgs versions of these models.

\subsection{A mass for the Standard Model photon}

As discussed in \cite{Reece}, experimental constraints on the mass of the photon are stringent (see \cite{Goldhaber:2008xy} for a review), with kinematic tests placing an upper bound $m_\gamma \lesssim 10^{-14} \ {\rm eV}$ \cite{Wu:2016brq,Bonetti:2016cpo,Bonetti:2017pym}.
From a theoretical perspective, engineering such a small, but non-zero, photon mass, while complying with quantum gravity constraints on effective field theories, may turn out to be challenging. Indeed, experimental constraints combined with the conjectures of \cite{Reece} provide a compelling argument that the photon must be massless.

The argument is roughly as follows: If the SM photon had a mass $m_\gamma \lesssim 10^{-14} \ {\rm eV}$ arising from Higgsing, this would imply a scalar mode at roughly the same scale, coupling to the photon with strength $e \approx 0.3$, a possibility which is clearly ruled out. A St\"uckelberg mass does not alleviate the problem in light of the conjectures of \cite{Reece}. If the radial mode conjecture applies, then the obstruction is precisely as in the Higgs case. Even ignoring the radial mode conjecture, the theory would still feature a cut-off at scale $\Lambda \sim \sqrt{f_\theta M_{Pl}} \sim {\rm MeV}$, which is also incompatible with experiment.

As noted in \cite{Reece}, a way out would be to assume that the electromagnetic gauge coupling is in fact quantized in units of $e_\gamma \ll e$, so that a unit-charge Higgs or St\"uckelberg axion could account for a non-zero SM photon mass, while remaining compatible with experimental constraints. For instance, for $e_\gamma \sim 10^{-14}$, and $v$ (or $f_\theta$) $\sim 1 \ {\rm eV}$, one would have $m_\gamma \lesssim 10^{-14} \ {\rm eV}$, whereas such a light scalar excitation with electric charge $\mathcal{O} (10^{-14})$ remains compatible with current measurements \cite{Davidson:2000hf,Vogel:2013raa}.

To some extent, the $e_\gamma \ll 1$ scenario proposed in \cite{Reece} can be realised through the broken clockwork constructions presented here.
Specifically, it corresponds to applying the clockwork mechanism as discussed in section~\ref{sec:review} to the hypercharge gauge group of the SM, and assuming the rest of the SM field content remains localized on the $j=0$ site. In this case the electromagnetic charge quantum would now be
\begin{equation}
	e_\gamma \sim \frac{e}{q^N} \ll 1 \ ,
\end{equation}
assuming that the vev of the Higgs (or the period of the St\"uckelberg axion) fields responsible for the clockwork mechanism is much larger than the scale of electroweak symmetry breaking in the SM.
In this construction, SM fermions carry electric charges in units of $q^N e_\gamma$.
Moreover, a unit charge Higgs (or St\"uckelberg axion) localized on the $j=N$ site will lead to a small mass for the photon, of order
\begin{equation}
	m_\gamma \sim e_\gamma v  \sim \frac{e v}{q^N} \ .
\end{equation}

In this construction, however, $v$ (or $f_\theta$) cannot be taken as low as $\sim 1 \ {\rm eV}$. As discussed in section~\ref{sec:vectorCWhiggs}, a distinctive feature of clockwork constructions is the presence of massive vector modes at scale $m_V \sim g q v \lesssim v$. Since these massive vectors are not strongly localized across the quiver, their coupling to localized matter will be $\mathcal{O}(1)$. Massive copies of neutral SM gauge bosons with $\mathcal{O}(1)$ couplings to SM fermions are highly constrained by experiments, with direct searches putting lower bounds on their masses of order $\sim 4 \ {\rm TeV}$ (see e.g.~\cite{Aaboud:2018bun}).
Constraints on clockwork models are likely to be stronger, given the multiplicity of states, so that $v$ (or $f_\theta$) $\gtrsim 10 \ {\rm TeV}$ to remain consistent with current data. 
Together with the experimental requirement $m_\gamma \lesssim 10^{-14} \ {\rm eV}$, generating a photon mass consistent with all constraints and the conjectures in \cite{Reece} is possible provided
\begin{equation}
	q^N \gtrsim 10^{26} \left( \frac{10^{-14} \ {\rm eV}}{m_\gamma} \right) \left( \frac{v}{10 \ {\rm TeV}} \right) \ ,
\end{equation}
which can be achieved, for instance, for $q = 5$ and $N \approx 38$.

\section{Conclusions}
\label{sec:conclusions}

Looking for consistency conditions applicable to effective field theories by demanding they remain compatible with a UV embedding into a theory of quantum gravity -- the philosophy behind the Swampland program -- is a promising enterprise. At a time of much needed guidance in Beyond-the-Standard-Model physics, given the lack of unambiguously positive results from collider and dark matter detection experiments, a well-defined set of conditions would be welcome. However, for the Swampland program to be relevant to the low energy theorist building models to solve the various problems that afflict particle physics today, the conjectures must be sufficiently compelling as applied to four-dimensional effective field theories in the far IR.
In particular, the existence of controlled counter-examples that apparently violate Swampland conjectures at low energies, but nevertheless allow for consistent (albeit partial) UV-completions,
weakens the extent to which the conjectures provide a meaningful guiding principle for IR phenomenology.

We have shown that the conditions proposed in \cite{Reece} (conjectures (\emph{1}) and (\emph{2}) in the Introduction), aimed at effective field theories containing light vectors with St\"uckelberg masses, can be parametrically violated while still remaining compatible with a UV-completion into a theory that satisfies them. Moreover, such UV-completion can be implemented both in the context of the Higgs and St\"uckelberg mechanisms.

The constructions we have considered here are a small modification of the original clockwork axion models \cite{Choi:2015fiu,Kaplan:2015fuy} as applied to vectors.
The axion constructions of \cite{Choi:2015fiu,Kaplan:2015fuy} were first proposed in order to obtain a low energy effective theory featuring an axion field with an effective decay constant $f \gg M_{Pl}$, but without introducing super-Planckian parameters from the UV perspective.
The vector version was first introduced in \cite{Saraswat:2016eaz}, in order to illustrate how versions of the WGC satisfied in the UV may be parametrically violated by the effective theory that remains at low energies.
It is of little surprise, then, that a small modification of this construction -- namely, through the breaking of the continuous symmetry that the clockwork mechanism leaves unbroken -- would serve to circumvent analogous versions of the Swampland conjectures that apply to theories with massive vectors.

An open question remains as to whether a continuum version of clockwork theories (both in their unbroken and broken versions) exists that exhibits the same properties with respect to WGC and more general Swampland arguments.
In line with the discussion in \cite{Craig:2017cda}, a continuum version of the standard clockwork constructions discussed here can be obtained by promoting the discrete lattice to a flat extra dimension, and including bulk and brane mass terms. Further breaking of the $U(1)$ gauge symmetry that remains unbroken from the four-dimensional perspective can be obtained, for example, through a Higgs field living on the opposite brane to which the massless field is exponentially localized.
It is unclear, however, the extent to which compactness and separation of scales with respect to WGC cut-offs are mimicked in this continuum version, leaving it a topic for future work.

\section*{Acknowledgments}
The research of IGG is funded by the Gordon and Betty Moore Foundation through Grant GBMF7392. The research of NC is supported in part by the US Department of Energy under the Early Career Award DE-SC0014129 and the Cottrell Scholar Program through the Research Corporation for Science Advancement. Research at KITP is supported in part by the National Science Foundation under Grant No.~NSF PHY-1748958.

\bibliography{massivephotonWGC_refs}

\end{document}